\documentclass{article}
\usepackage{amsfonts,natbib}

\title{A Role for the Fauxrizon in the Semiclassical Limit of a Fuzzball}
\author{Mike D. Schneider\footnote{Department of Philosophy, University of Missouri} \footnote{I thank Samir Mathur, Nick Huggett, Eugene Chua, and Diana Taschetto for early discussions, and the latter three I also thank for early draft feedback. I also thank my audiences at the 2022 German Physical Society Annual Conference and at the Foundations Seminar in the Black Hole Initiative at Harvard University during Fall 2023. Finally, I am indebted to my anonymous reviewers for their helpful critiques. A good deal of the work that went into this article was completed during the 2021-2022 academic year while I was a postdoc on the Beyond Spacetime project at University of Illinois Chicago, funding provided by the John Templeton Foundation grant no. 61387. I am grateful to the South Shore Line, on whose commuter trains I wrote (nearly all of) the first draft.}}
\date{}

\begin{document}

\maketitle

\abstract{Recent work on the status of astrophysical modeling in the wake of quantum gravity indicates that a `fauxrizon' (portmanteau of `faux horizon'), such as is relevant to understanding astrophysical black holes according to the fuzzball proposal within string theory, might ultimately solve the familiar black hole evaporation paradox. I clarify, with general upshots for the foundations of quantum gravity research, some of what this suggestion would amount to: identification of intertheoretic constraints on global spacetime structure in (observer-relative) semiclassical models of fuzzballs.}

\section{Introduction}

\citet{manchak2018information} have argued that the familiar black hole `evaporation' paradox in semiclassical gravity may be presented succinctly as three ``well-motivated and widely accepted assertions'', which jointly entail contradiction. (Why does this suffice for paradox? See footnote 2 in their article, with reference to existing philosophical literature on the notion.) The upshot: ``[...] any `solution' to the paradox will require one to reject one of these assertions'' [p. 612]. Unlike the closely related `Page-time' paradox in black hole thermodynamics (cf. discussion in \citep{wallace2017black}), the assertions relevant to the evaporation paradox, as \citeauthor{manchak2018information} identify, exclusively concern assumptions about global spacetime structure. Talk of solving the evaporation paradox therefore amounts to talk of some new lesson about global spacetime structure, were we so lucky as to find cause --- new, satisfying theory --- to leave our current (semiclassical, paradoxical) understanding of black holes behind.

A point of clarification. As it happens, those engaged with paradox in contemporary black hole research are, arguably, not foremost preoccupied with learning about global spacetime structure. Primarily, black hole research in quantum gravity focuses on the subject of the physical fate of (quantum) information through the total lifetime of a black hole, and this research focus would seem to spotlight means by which we might solve other named instances of paradox, e.g. the Page-time paradox, whose explicit formulations place them nearer to that subject. But paradox is paradox, simply put; making it explicit in certain terms (say, those that enter into a formulation of the Page-time paradox) for the sake of certain research projects does not undermine the use of many other terms (say, those exclusively involving global spacetime structure) that others might see fit to call upon in a different act of formulating, perhaps for different research ends. There is more than one lesson we stand to learn about black holes, given our current (semiclassical, paradoxical) understanding of them. And there is value in remarking on potential lessons that lurk outside of the research spotlight.  

Recently, \citet{huggett2021lost} have claimed that, within a string theory approach to quantum gravity research, the fuzzball proposal developed by \citet{mathur2005fuzzball,mathur2012black} avoids the constellation of worries in astrophysical modeling that surround the evaporation paradox. \citeauthor{huggett2021lost}'s contribution on this particular topic (within the context of a much larger foundational argument about the fungibility of spacetime geometry in string theory) is essentially to direct attention to some interesting new quasi-local physics that would be relevant to a stringy astrophysical black hole that is modeled, accordingly, as a fuzzball. They dub the interesting new physics the `fauxrizon', in a nod to its uncanny likeness to the horizon in more familiar cases of semiclassical black holes. Namely, like the horizon in the familiar cases, the fauxrizon marks ``the end of space external to the black hole'' \citep[p. 15]{huggett2021lost}. But unlike the horizon, beyond the fauxrizon, (effective) spacetime structure immediately fades out. Instead of admitting some description as a spacetime, the black hole interior in the fuzzball construction is rather described in terms of an irreducibly ``nonspatial, fundamentally stringy state'' [p. 15]. And note: implicit in this `effective spacetime up to the fauxrizon' characterization of the astrophysical black hole in string theoretic modeling, the relevant observer of the fuzzball is understood to be situated somewhere out in the effective exterior spacetime. (This detail will be important below.)  

In virtue of the above theoretical observation about stringy astrophysical black holes when modeled as fuzzballs, a natural impulse is to regard the familiar evaporation paradox as solved in a string theory approach to quantum gravity research, given the fuzzball proposal therein (that is, as the latter is advertised by its proponents). If that impulse proves just, so much the better for the fuzzball proposal; on a fairly standard view of progress in theoretical science, solving outstanding paradoxes is exactly what one looks forward to being accomplished by new proposals in research. Indeed, beyond just enthusiasm for the fuzzball proposal within string theory, a promised solution to the evaporation paradox could even provide extrinsic justification to continue pursuing a string theory approach to quantum gravity research in general, conditional on a newfound expectation that we may enjoy, at the end of the process of inquiry, those lessons as are now anticipated of the fuzzball proposal within the approach. So there are stakes in assessing the relevant claim.

Here, I specifically discuss the form of the inference from the identification of the fauxrizon to the conclusion that the fuzzball proposal in string theory solves the evaporation paradox --- particularly given \citeauthor{manchak2018information}'s well-placed emphasis on global spacetime structure in the presentation of the latter. The primary complication involved is that fuzzballs are not themselves well defined in the semiclassical limit. Hence, what I will ultimately argue is that, at present, the identification of the fauxrizon within the fuzzball construction is insufficient to regard the paradox as solved in a string theory approach to quantum gravity research (i.e. even granting the success of the fuzzball proposal, therein). On the other hand, its identification does helpfully reshape how we might eventually come to solve the paradox --- albeit by further technical means, which are presently obscure. 

Despite the details involved, I take the present discussion to be of some general import in the foundations of quantum gravity research, beyond just work on the fuzzball proposal in string theory (or on quantum gravity firewalls, slightly more generally --- cf. footnote \ref{fnfirewarll} below). A large part of the mystery in fundamental physics surrounding astrophysical black holes is the (non-effective) emergence of some non-trivial, classical global structure from an underlying quantized theory, given the relevant astrophysical modeling context (where, following the leads of \citet{stein1995some}, \citet{curiel2019schematizing}, and \citet{smeenk2020some}, implicit in that modeling context is a certain choice of how to \emph{schematize the observer}: in our case, as situated spatiotemporally to the exterior of the astrophysical black hole, suitably far away). Yet, this global structure is what is needed to actually solve, in accordance with the astrophysical application (and, hence, with the relevant empirical data, given the choice of observer schematic), the local geometrodynamics that arise in the semiclassical theory as an effective description of the dynamics within that same underlying quantized theory. Along these lines, discussing how the fauxrizon of a fuzzball might ultimately solve the evaporation paradox spotlights the peculiar interplay between boundary and bulk in the semiclassical limit of quantum gravity, specifically when theorizing about `strong-field' gravitational modeling contexts relative to `weak-field' observers placed out at a distance.

\section{The evaporation paradox}

In this section, I elaborate on the three assumptions about global spacetime structure in semiclassical gravity that jointly comprise the evaporation paradox, as presented by \citet{manchak2018information}. But first, a caveat. \citeauthor{manchak2018information} discuss these assumptions in terms of what is `physically reasonable' as a global solution to the dynamical equation that locally characterizes general relativity (GR), our current theory of gravity, as a classical relativistic field theory that may be understood to govern local spacetime geometry. This is a standard framing device in the philosophical literature on the physical foundations of (relativistic) spacetime. Yet, there is some awkwardness in this framing: most immediate, as has already come up, the evaporation paradox concerns semiclassical gravity, rather than the classical GR.

Moreover, even ignoring the distinction between semiclassical gravity and GR, this framing can lead to convoluted statements when discussing the representational capacities of spacetime models relative to a choice of observer schematic --- a subject that will be centrally important below. As \citet[p. 231]{fletcher2020representational} discusses, \begin{quote}[...] the representational capacities of a mathematical model depend not just on the particular set-theoretic object that constitutes it, but also how its users consider or intend it to be part of a larger class--e.g., that a particular spacetime model represents spacetime \emph{as} a Lorentzian manifold.\end{quote} In the wider context of the quotation, the intention of users to represent spacetime \emph{as} a Lorentzian manifold in the example (emphasis in the original) is presumably warranted by the users' specific embrace of the theory GR. But the role of the intentions of the users in this account has further consequences in our thinking about \emph{applications of the theory by those users to particular modeling contexts}. For instance, as discussed later in \citeauthor{fletcher2020representational}'s article, the representational capacities of a Schwarzchild spacetime of a given Schwarzchild radius (amongst the class of Lorentzian manifolds, equipped with its canonical standard of mathematical equivalence of models) include more than that which is strictly necessary for our applying GR to adequately model an arbitrary isolated Schwarzchild black hole (i.e. modeling it \emph{as} a Lorentzian manifold, per dictum of the theory). Yet, meanwhile, in other modeling contexts where units are externally fixed, the same spacetime model is rendered uniquely adequate (up to the canonical standard of mathematical equivalence of models relevant in GR) for accomplishing a very similar task, through application of the very same theory.

An upshot is that what it means for a spacetime model with various formal properties to be physically reasonable (or not) in a theory like GR is sensitive, case by case, to extra-theoretical details about the physical systems we happen to be interested in modeling according to the theory. Which physical systems we consider to be adequately modeled by means of the theory may therefore influence, case by case, what we deem physically reasonable in the theory. So, for instance, whether Schwarzchild spacetimes of sufficiently microscopic Schwarzchild radii are physically reasonable or unreasonable in semiclassical gravity will depend on context of application: modeling an isolated astrophysical black hole with a re-scaling of units --- reasonable; modeling with units fixed relative to other astronomical bodies of interest (or, say, to a fixed length scale that separates the target system from the relevant observer) --- not reasonable (as one would expect any sufficiently microscopic black holes to evaporate too rapidly to physically countenance, due to physical processes documented coincidentally below).

So, in light of this contingent character of the notion of physical reasonableness in the context of modeling, and also in light of the conceptual distance between GR and semiclassical gravity, I propose to amend \citeauthor{manchak2018information}'s treatment of the evaporation paradox (albeit only slightly). Namely: the three assumptions concern \emph{what it is descriptively adequate to say about} global spacetime structure\emph{ (i.e. as a matter of representation), specifically regarding applications of semiclassical gravity to the dynamical modeling of targets in astrophysics}. 

Here are the assumptions, so amended (original language in footnotes):

\begin{enumerate}  
    \item \emph{Lorentzian Manifold} (LM): Spacetime, in applications of the semiclassical theory, is represented by a (smooth) Lorentzian manifold without boundary.\footnote{``Any physically reasonable relativistic spacetime may be represented by a smooth (Hausdorff, paracompact) manifold without boundary, with a Lorentz signature metric'' [p. 618].}
    \item \emph{Global Hyperbolicity} (GH): Spacetime is necessarily globally hyperbolic in such applications.\footnote{``All physically reasonable relativistic spacetimes are globally hyperbolic'' [p. 614]. For reasons that will become apparent below, what I take to be physically interesting about GH in the context of semiclassical gravity is what follows from its pairing with LM. As such, I have in mind by GH the standard technical definition of global hyperbolicity in the foundations of spacetime literature, which indeed presupposes LM. But see footnote \ref{fnglobalhyperbolicity} below on this topic, which includes reference to a discussion by \citeauthor{manchak2018information} about troubles in any search for a technical condition sufficiently `like' global hyperbolicity, which would be required in order to formulate a version of GH that may stand independent of the status of LM.}
    \item \emph{Evaporation Geometry} (EG): Astrophysical black holes can, when adequately represented as a spacetime in the semiclassical theory, include a (non-trivial) horizon and nonetheless fully evaporate.\footnote{\label{fnNaked}``Some evaporation spacetimes are physically reasonable'' [p. 616]. Note that \citeauthor{manchak2018information} do not here use the term `naked singularity'. But they do identify the previous assertion with the `cosmic censorship' hypothesis, which (as they note) is usually understood as the hypothesis that naked singularities are impossible. Meanwhile, the result they employ to argue, given LM, that GH and EG jointly entail contradiction is due to \citet{kodama1979inevitability}, as appears in an article whose title employs the term. An earlier version of the present manuscript expressed EG in terms of a naked singularity. However, in deference to reviewers' concerns that the term `naked singularity' might sound too benign, this framing has since been revised. Regardless of terms employed, the key point of EG is, first, that black holes evaporate. But then, in so doing, there comes to be a kind of peculiar geometric structure generally associated with the full lifetime or four-dimensional, global spacetime description of that evaporation process. This peculiar geometric structure is what \citeauthor{manchak2018information} capture in terms of their explicit definition of the class of `evaporation spacetimes', and is also what I have tried to capture informally in my own statement of EG in the main text.}
\end{enumerate}

In a moment, I will discuss each of these assumptions, including their meanings and motivations, at length. For now, it is sufficient to note that the three assumptions are each widely accepted as well-motivated, and EG in particular is understood as a consequence of physical Hawking radiation thought to be emitted by astrophysical black holes, treated semiclassically. Yet, given LM, GH precludes the peculiar global geometric structure involved in EG --- hence, a paradox. 

The amendments to the three assumptions have another advantage: they make clear the evaporation paradox's underlying modal character. In \S\ref{secFuzz} below, this will be important. Namely, GH and EG are respectively claims about necessary and possible features of adequate choices of representation of astrophysical black holes as spacetime models within the context of semiclassical gravity, given the formal characterization of spacetimes in semiclassical gravity supplied by LM. But in conjunction with that formal characterization supplied by LM, GH implies that there \emph{cannot be} spacetime geometry of the kind involved in EG in adequate representations of astrophysical black holes as spacetime models within semiclassical gravity, whereas EG insists that there \emph{can be}. 

Finally, notably left out of the paradox is an assumption, alongside the `without boundary' stipulation in LM, that spacetimes in semiclassical gravity are inextendible. This omission is important: in \S\ref{secFuzz}, such an assumption will be violated. Yet, it is arguably at odds with standard attitudes in the foundations of spacetime literature to allow for extendible spacetimes in applications of relevant theories. For ease of mind on this matter, see \citep[ch. 6]{manchak2020global} for some critical discussion on the logic of extendibility, which counsels against standard attitudes.

\subsubsection*{LM Assumption} 
LM articulates the usual broadest constraints in GR on the kinematics of classical gravity.\footnote{See, e.g., discussions in \citep{hawking1973large}, \citep{geroch1979global}, and \citep{manchak2020global}.} What might these have to do with applications of semiclassical gravity, in the course of quantum gravity research? As I discuss in \citep{schneider2020s}, given our current physics that includes GR, modeling in semiclassical gravity is intended to provide  ``claims about what we might infer from contemporary physics about approximations of future physics'' [p. 10]. In light of this pragmatic view of the applications of semiclassical gravity, it is reasonable to assume that the broadest constraints from our current best classical theory are carried over into the semiclassical. 

On the other hand, the relevance of the future theory of quantum gravity to semiclassical modeling on this pragmatic view also means that independent reasons might be supplied, which ultimately counsel in favor of kinematic constraints that are at odds with those familiar from the classical theory. But such independent reasons, supplied in the course of quantum gravity research, are grounded in relationships between our current physics and the future theory yet to be developed. This would suggest that, unlike the constraints carried over from GR, any other constraints would have the status of speculations in the course of quantum gravity research.

So, consider \citeauthor{manchak2018information}'s [p. 618-619] objection to \citet{maudlin2017information}, who the former interpret as rejecting LM:\footnote{\label{fnglobalhyperbolicity}Specifically, \citeauthor{manchak2018information} quote \citeauthor{maudlin2017information}'s assertion that ``our evaporating black hole space-time is not a manifold'' as indication that \citeauthor{maudlin2017information} rejects a ``third, suppressed premise'' [p. 618] of the paradox --- namely, the one I have named LM. As \citeauthor{manchak2018information} stress, \citeauthor{maudlin2017information} does not reject GH or EG: ``The idea is that there is some critical surface, $\Sigma_{crit}$,
after which Cauchy evolution proceeds along \emph{disconnected} Cauchy surfaces. There is no failure of determinism, or of unitarity; rather, one just needs to take account of
the part of the global state corresponding to $\Sigma_2$ that remains trapped behind the event horizon'' [p. 618]. Here, crucially, $\Sigma_{crit}$ passes through an `evaporation event' that renders \citeauthor{maudlin2017information}'s proposed spacetime in tension with LM even as it preserves GH and EG, which is just the point \citeauthor{manchak2018information} go on to emphasize. On the other hand, as they ultimately conclude, abandoning LM leaves things admittedly vague regarding the continued --- now independent --- articulation and justification of GH: ``It follows that even if one can make sense of a notion of `global hyperbolicity' that includes the spacetime Maudlin proposes (with evaporation event), it is not clear what significance that has [....] But we think it is valuable to reflect on just what is being given up'' [p. 624]. One thing given up, they show, is the sense in which global hyperbolicity precludes such a spacetime as Maudlin proposes from having ``really big gaps'' [p. 626].}\begin{quotation} We do not deny that there may be good reasons to reject [LM]. After all, the arguments in favor of [EG] rely on a semi-classical analysis according to which spacetime is treated classically and the radiation is treated quantum field theoretically. One might well expect that a full understanding of Hawking radiation will wait on a theory of quantum gravity---and on several approaches to quantum gravity, the description of space and time as a smooth manifold breaks down [....] But whatever else is the case, observing that [LM] may be rejected does not simply dissolve the paradox.\end{quotation} There are two points to emphasize here. First, it seems right to say that physicists involved in quantum gravity research are often free to speculate that the development of the future theory may ultimately supply resources relevant to interpreting our current physics, which were not initially available in consultation of our current physics alone. On such a view, \citeauthor{manchak2018information} may here be understood as outlining a common speculation about what will come of quantum gravity research: that there may eventually be reason supplied by the future theory to justify our deviating from LM when modeling what we are today comfortable recognizing as astrophysical black holes. 

Second, it is crucial that, \emph{nonetheless}, deviating from LM in our modeling astrophysical black holes in light of quantum gravity, even as a matter of speculation, is not what would solve the evaporation paradox. To solve the paradox, it must be that the speculation would, moreover, compel us to change how we see fit to model astrophysical black holes \emph{when we continue to have cause to do so semiclassically}. That is, something about the modified description of an astrophysical black hole provided by the future theory of quantum gravity must, consequent to our embrace of that future theory, teach us something new about what amounts to an adequate description of the modeling target in a suitable semiclassical limit. 

But what it might teach us --- and how --- is hard to say in advance. For instance: arguably, the fuzzball proposal in string theory, discussed below, presents one case of what \citeauthor{manchak2018information} envision, wherein the description of space and time as a smooth manifold breaks down at a `fauxrizon' relevant to the quantum gravity treatment of an astrophysical black hole. (As \citet [p. 14]{huggett2021lost} observe of the fuzzball proposal: ``certainly it falls into the category of `drama at the horizon' '' --- whereas the classical GR tolerates no drama.) Yet, in this case, LM appears to remain entirely intact, so far as concerns the \emph{semiclassical description of that black hole}. Namely, as I will argue following \citet{huggett2021lost}, it is plausible that adequate semiclassical descriptions of the astrophysical black hole relative to an exterior `weak-field' observer, in light of the identification of the fauxrizon in the fuzzball construction, are provided by exterior black hole spacetimes. So if the breakdown of space and time as a smooth manifold at the fauxrizon is to teach us how the paradox is solved, it must be that the fuzzball construction within string theory, in virtue of that breakdown at the fauxrizon, gives us further reason to reject either GH or EG in the semiclassical limit. 

\subsubsection*{GH Assumption}
GH ensures the predictability and retrodictability of the evolution of classical fields in the spacetime, including the metric itself when understood locally (i.e. as a dynamical field on some underlying smooth manifold, with distinguished hypersurfaces identified as Cauchy). As \citeauthor{manchak2018information} note, this is a version of the cosmic censorship hypothesis originally due to Penrose. Also as they note, at least in the context of the classical GR, predictability and retrodictability provide a neat (classical) description of that which is at stake in standard talk about the conservation of `information' in contemporary theoretical physics (see also \citep[\S 2]{wuthrich2017black}). 

Understanding cosmic censorship in terms of maximal domains of predictability and retrodictability follows a tradition in the philosophical foundations of GR inaugurated by \citet{earman1995bangs}. \citeauthor{earman1995bangs} considers this topic as closely allied with various commitments that classical relativistic field theories be deterministic. And although global hyperbolicity is not necessary for determinism in these theories, it does seem sufficient. This is (likely) because global hyperbolicity implies that the spacetime, per LM, is diffeomorphic to $\mathbb{R}\times S$, with $S$ a smooth spacelike Cauchy surface \citep{bernal2003smooth}. Consequently, the space of suitable initial data on $S$ for a relativistic field can be arranged in a one-to-one relationship with global field configurations that are consistent, given a local dynamics of the field, with the maximal evolution of that initial data off of any one such embedded hypersurface.\footnote{This exposition should highlight one sense in which global hyperbolicity is not necessary for determinism. For more, see \citep{friedman2004cauchy}.} 

Two remarks are, at this point, in order. First, I have just included a caveat that global hyperbolicity is not necessary for determinism. In \citep{lesourd2018causal}, a theorem is provided in support of the view that fully evaporating black holes imply a failure of causal continuity, above and beyond a failure of global hyperbolicity. Accompanying that result is a discussion of the relationship between causal continuity and predictability/determinism, which runs more or less the same as that just discussed (except that causal continuity does not imply a product structure $\mathbb{R}\times S$, with $S$ a smooth spacelike Cauchy surface). As stated within that discussion, causal continuity has historically been proposed as a weaker condition than global hyperbolicity that is nonetheless sufficient to ensure predictability. Meanwhile, it is suspected that causal continuity is necessary for predictability, though the question remains open. If causal continuity does prove necessary, then the theorem would seem to secure the sense in which GH, indeed via a relaxation thereof, fits together with LM and EG in the manufacture of paradox concerning determinism in full black hole evaporation, or (in turn) paradox concerning the conservation of information through the same.

Second, all of the above is a matter of classical GR. In the context of semiclassical gravity, whether one focuses on causal continuity or global hyperbolicity with regards to the causal structure of the spacetime, there is a further question about how we might proceed to think about the quantum fields one ought to be prepared to define there. And, although \citeauthor{manchak2018information} do not stress this point in their discussion of GH, it is the product structure $\mathbb{R}\times S$ ensured in the case of global hyperbolicity that one may exploit, in the context of quantizing what is otherwise a classical relativistic field theory defined locally on the spacetime. Namely: the association of classical states of the field with their immediate, quasi-local evolution off of a particular Cauchy hypersurface embedded in the spacetime allows for an unambiguous means of constructing a corresponding quantum field theory on that spacetime \citep{wald1994quantum}. There may, of course, turn out to be other ways to proceed in quantization if the spacetime is not globally hyperbolic. But there would appear to be significant costs \citep{yurtsever1994algebraic,friedman1997field}. Along these lines, it is worth noting that global hyperbolicity remains as a constraint on global spacetime structure, even in much more recent efforts to define `locally covariant' formulations of quantum field theory --- that is, formulations that are `background independent' in a sense licensing talk about `quantum fields', analogous to classical fields, as physical fields that may be relocated between arbitrary (allowed) spacetime backgrounds \citep{redei2014categorial,brunetti2016quantum}.\footnote{\label{fn-nonGH}That being said, efforts are ongoing to construct quantum field theories on various types of non-globally hyperbolic spacetimes, including those with timelike boundaries \citep{benini2018algebraic} or, indeed as are relevant here, for evaporating spacetimes and the case of spatial topology change \citep{janssen2022quantum}. (I am grateful to an anonymous reviewer for this final reference.)} 

This formal state of affairs is invaluable for delimiting semiclassical gravity. After all, semiclassical gravity is concerned with the gravitational self-interaction of material systems, precisely when the quantum descriptions accorded to those systems, i.e. by an appropriately quantized theory, cannot be ignored. (Though, when quantum fluctuations are large in that description --- such as in the case of macroscopic superpositions of matter --- semiclassical gravity too is inadequate.) Together with LM, GH thereby ensures that we at least know how, in the first place, to associate the spacetime backgrounds for relativistic quantum field theories with arbitrary \emph{allowed} gravitational field states. This paves the way for a perturbative approach to studying the back-reaction of the quantized fields on such gravitational field states, with the latter now treated semiclassically: as a spacetime expression of the mean-field effects relevant to a gravitationally-coupled quantum system, which is otherwise modeled by a state within the (non- gravitationally coupled) quantum field theory defined on top.

\subsubsection*{EG Assumption}

EG concerns the dynamical process of semiclassical black hole evaporation, relative to an exterior `weak-field' observer: an observer who is schematized in the semiclassical theory as a maximal wordline in spacetime sufficiently far away from the strong-field gravitational system --- i.e. the black hole interior --- as to be adequately regarded as `near' null infinity, for all proper time. Assume, for simplicity, the case of black hole spacetimes with eventually zero angular momentum and charge. As originally noted by Hawking, a quantized scalar field in its vacuum state associated with any such black hole spacetime background is thermal at future null infinity, i.e. in the eventual `vicinity' of the observer. Although vacuum states are globally defined and utterly non-localizable in spacetime, mode-by-mode conservation arguments may be used to motivate the expectation that these thermal modes are dynamically produced by the black hole, to then escape to future infinity. This dynamical production process is called Hawking radiation. And since the Hawking radiation modes indeed manage to escape, one associates the origins of that radiation with physical processes that are restricted quasi-locally to a neighborhood of the event horizon of the black hole, according to the semiclassical description. Famously, this is a locus of concern in our semiclassical means of modeling black holes, in the course of quantum gravity research: according to classical GR, there should be nothing whatsoever that is significant about the local physics present along that horizon --- i.e. no `drama'. Yet, merely by treating matter as a quantum system, vacuum interactions between that matter and the black hole background imply novel physics in the vicinity.

Pushing onward nonetheless, it is natural to try to model the back-reaction of Hawking radiation on the background gravitational field state. One thereby expects from semiclassical gravity dynamical loss in the black hole mass parameter,\footnote{Though, see \citep[\S 2]{belot1999hawking} for some qualification about the nature of this expectation.} and with that loss the event horizon dynamically shrinks (as assessed in terms of suitably chosen hypersurfaces, which foliate the neighborhood of the relevant exterior weak-field observer). Eventually, one expects the black hole to fully evaporate (relative to that observer). But further details will not be important here; EG is a statement about the peculiar global spacetime structure where such a complete evaporation process occurs. 

Understandably, EG is where lies all substantive controversy about the status of the evaporation paradox in contemporary quantum gravity research. Namely, there is a sense in which this dynamical process simply lies beyond the scope of semiclassical gravity, so that it may be disregarded in the course of our efforts to infer from contemporary physics about approximations of future physics. Recall from the discussion of GH that the scope of semiclassical gravity is delimited by our knowing how to quantize field theories in the appropriate spacetime setting. If our understanding of the dynamical process of black hole evaporation, per EG, implies that GH is false, we therefore, arguably, have a breakdown of the applicability of the semiclassical theory to our modeling astrophysical black holes. So there is some cause to be skeptical that EG arises within applications of semiclassical gravity, carefully construed.

Indeed, one may suspect that semiclassical evaporation dynamics fail to be metastable as a physical description of astrophysical black holes in semiclassical gravity, in the course of quantum gravity research. Intuitively: we were led to identify evaporation as a dynamical process in semiclassical gravity because of radiation modes associated quasi-locally with the event horizon of the black hole, whereas in the black hole's final moments, when the mass parameter runs down to $0$, the event horizon abruptly vanishes. Or, put more formally: one reason to doubt the metastability of semiclassical black hole evaporation over the lifetime of an astrophysical black hole is due to the appearance of singular limits as the mass parameter runs down to $0$ in a finite time, in a variety of equations that are relevant in motivating the evaporation dynamics in the first place. 

On the other hand, I take it that this is just another way of framing that there is indeed a paradox here, upon one's embrace of LM, GH, and EG in the context of semiclassical modeling. Skepticism specifically focused on EG is then an expression of the kind of speculation sketched above: that we may eventually find cause in the future quantum gravity theory to reject specifically EG in the semiclassical limit, thereby solving the paradox. For instance, \citet{belot1999hawking} note the possibility of a ``thunderbolt evaporation'' as one avenue of escaping the paradox. In this case, one readily accepts LM and GH, and then hopes that, in virtue of some further reason to come (i.e. as a lesson from future physics), the process of complete semiclassical black hole evaporation --- even as just described --- simply does not entail the peculiar global spacetime structure otherwise claimed in EG, instead singling out some other global geometry that is not so causally pathological in the face of GH. Alternatively, one might hope that, in virtue of some further reason to come, the process of semiclassical black hole evaporation is simply incomplete, entailing a black hole remnant of some non-trivial mass \citep{bokulich2001black}. Or perhaps there may turn out to be a quantum gravity `bounce', whereby shrinking black holes eventually transition into white holes --- at a minor cost of apparently violating familiar energy conditions in the semiclassical description of that new physics \citep{malafarina2017classical}.\footnote{These hopes are intended to be illustrative of the general point. Recalling remarks in the Introduction, I am here ignoring further worries stemming from consideration of the Page-time paradox (which could strike against hope in some of these instances).}

All of this is to say: set aside reservations that black hole evaporation may simply be a dynamical process beyond the scope of semiclassical gravity. Except as a matter of speculation about the future theory of quantum gravity, such a view is not dispositive of rejecting EG in particular as a solution to the paradox today.  

\section{\label{secFuzz}Fuzzballs}

In a string theory approach to quantum gravity research, \citet{mathur2005fuzzball,mathur2012black} has argued for a `fuzzball' understanding of stringy astrophysical black holes. As helpfully discussed by \citet{huggett2021lost}, this picture replaces the event horizon of any such black hole with (what they dub) a `fauxrizon', at which spacetime immediately fades out as an effective description of the physics, in favor of an irreducibly stringy description of the black hole interior as full of quantum hair [p. 14]. Importantly, such a fuzzball would seem to be (by hypothesis) an unproblematic stringy construction in the future theory of quantum gravity. As \citeauthor{huggett2021lost} summarize [p. 15]:\begin{quotation} [...] we have an example where it is obviously inappropriate to ascribe a classical geometry to the interior, along the lines suggested by Polchinski earlier. Clearly in this case, black holes have rather profound implications for the nature of spacetime! Moreover, Maudlin’s construction\footnote{The reference here is to the same article by \citet{maudlin2017information} as is discussed at length by \citet{manchak2018information}. (\citeauthor{maudlin2017information}'s construction within that article is discussed obliquely in footnote \ref{fnglobalhyperbolicity} above.) The previous reference to Polchinski is \citep{polchinski2017black}.} again does not apply; unitarity – and indeed information conservation – is obtained by the details of the fuzzball dynamics.\end{quotation} 

Glossing over the details of their remarks, it is plausible that nothing troubling remains of astrophysical black holes, when understood as fuzzballs (or at least, nothing that we are prepared by black hole evaporation to note as troublesome!). In particular, work in the fuzzball proposal is primarily focused on recovering familiar black hole evaporation phenomena in a thermodynamic description of the relevant fundamental physics, where the underlying unitary machinery of string theory promises to dispel all mystery about information loss as arises in the semiclassical treatment. Meanwhile, there is some interesting new quasi-local physics that we have learned about astrophysical black holes, treated accordingly --- namely, what \citeauthor{huggett2021lost} have dubbed the `fauxrizon', which falls immediately beyond a region of effective physics adequately described in geometric terms as an exterior black hole spacetime. (Why not treat the fauxrizon as coincident in spacetime with a traditional horizon, or else as coincident with a timelike membrane situated just before there? As \citeauthor{manchak2018information} caution in general, one might be wary of thinking about such novel, exotic constructions as being co-located with a manifold boundary, such as might then be embedded in a larger manifold --- e.g. the horizon within the complete black hole spacetime that includes an interior. Manifold boundaries are necessarily topological manifolds of $n-1$ dimensions for an $n$ dimensional spacetime, and such a demand on topology may very well be inappropriate in the given case.)

But are all worries about black hole evaporation really obviated by the proposal? In particular: what becomes of the \emph{evaporation paradox}? As I will argue, the answer is subtle; not enough is yet known about fuzzballs to neatly state what stands to be learned about global spacetime structure in our understanding of black holes, given the current promise of fuzzballs.

\subsection{Global structure in the aftermath of the fuzzball proposal}

To begin, note that the fuzzball construction as sketched is irrelevant to the three assumptions that comprised the evaporation paradox, since the latter, as stated, were restricted in focus to applications of semiclassical gravity. Here, by contrast, we are considering models of the same astrophysical systems, and presumably relative to the same observers, only now in the context of string theory. On the other hand, the fuzzball construction as sketched would become relevant to the three assumptions, were we to equate the stringy model, intertheoretically, to some description in a suitable `semiclassical limit': in particular, as relevant for some observer localized as a quantum clock in effective spacetime suitably far away from the fauxrizon so as to correspond, in the semiclassical theory, with the schematic of the exterior weak-field observer outlined above. In that case, since the fauxrizon quarantines the irreducibly stringy quantum hair from the effective spacetime region of the fuzzball that is coincident with the exterior of a semiclassical black hole, what is recovered in the semiclassical theory (relative to the chosen observer) is exactly the latter: an exterior black hole spacetime. This is an extendible spacetime that comprises only the region (without boundary) of a black hole spacetime that is retrodictable from future infinity. On this account, the fauxrizon itself would simply disappear, along with all the stringy quantum hair `beyond' it, in the course of taking the relevant limit.\footnote{\label{fnfirewarll}There is a sense in which the fuzzball proposal may amount to one explicit realization of a more general speculation about the presence of a quantum gravity `firewall', which replaces the horizon in a future quantum gravity theory and is (therefore) sometimes discussed in terms of a `drama at the horizon' solution to the evaporation paradox. Namely, as \citet[p. 13]{huggett2021lost} note, what it means to accept a `drama at the horizon' solution to the evaporation paradox could include accepting ``even the absence of a horizon in the first place. Objects -- and observers -- never really pass the horizon [...] there only is the exterior description, no complementary description according to an infalling observer. In this case, it has been suggested [by \citet{polchinski2017black}] that there is no classical spacetime interior either [....]'' Throughout this paper, I focus on the fauxrizon of a stringy fuzzball for definiteness. But it may well be that the inference from the identification of the fauxrizon to the conclusion that the fuzzball proposal solves the evaporation paradox is just one specific realization of an inference from the identification of a quantum gravity firewall to the conclusion that it is in virtue of our discovering the firewall in the course of quantum gravity research that the evaporation paradox is solved. If this is correct, my argument throughout the paper about what more exactly is needed to warrant the inference in the case of the fuzzball proposal applies equally to a more general study of quantum gravity firewalls.}

Exterior black hole spacetimes readily satisfy LM, and we may just as well stipulate that they satisfy GH. That is, in the previous paragraph, exterior black hole spacetimes are presented in terms of retrodiction from future infinity. Although vaguely specified, this immediately ensures LM. Meanwhile, if, consistent with this vague presentation, we understand our models of astrophysical black holes in the semiclassical theory to idealize the relevant astrophysical systems as isolated strong-field gravitational systems in asymptotically Minkowski spacetime, GH follows from properties of the latter.\footnote{It is an interesting question under what circumstances this idealization remains apt. It may be, e.g. in light of global asymptotic rotation or --- more realistically --- a positive cosmological constant that endows all observers with non-trivial cosmological horizons, that sufficiently distant observers of a strong-field gravitational system are in some cases \emph{too far away} to be aptly regarded as residing in a `weak-field' regime while performing their observations.  Might GH in the semiclassical limit be an observer-relative feature of the fuzzball, mixed up in the recovery of the familiar schematic of the exterior `weak-field' observer?}

In fact, it may be that GH is ultimately difficult to secure, as a matter of stipulation.\footnote{I thank my anonymous reviewers for pressing this point.} While exterior spacetimes for non-evaporating black holes can easily be globally hyperbolic, it is trickier to imagine what might ever suffice as a choice of Cauchy surface for an exterior portion of an evaporating black hole spacetime. In the absence of explicit argument for GH, it may be that fuzzballs would force us to confront failures of global hyperbolicity in the semiclassical limit. 

Accepting such a failure of strong cosmic censorship in the semiclassical limit may or may not be alarming (see footnote \ref{fn-nonGH} and surrounding discussion). Still, it seems to me that the promise of the fauxrizon (to the extent that one is optimistic) is that stringy physics will ultimately prove to be quarantined in black hole scenarios, relative to a distant exterior observer, from an otherwise mundane exterior spacetime in which they reside. Unitarity in the underlying string theory being associated, loosely, with a globally hyperbolic target spacetime, that promised fact of quarantining would seem to point to a globally hyperbolic exterior portion. On the other hand, thinking of the fauxrizon instead as a more sophisticated description of a timelike quantum membrane counsels in the other direction, so intuitions are mixed.

In the face of ambiguity, I suggest a choice made for definiteness in the argument that will follow: suppose that GH, like LM, is seen as reasonable in descriptions of a fuzzball obtained in the semiclassical limit, relative to a suitable choice of exterior weak-field observer. The upshot is that, in order to ultimately solve the evaporation paradox, something about the fauxrizon must supply a reason to discard EG.\footnote{\label{fn:loophole}Note that the sidelined concerns about the fate of GH function here as possible loopholes.} Does it? Here the subject gets muddy. It is (plausibly) true that the fauxrizon in the full quantum gravity description undermines the descriptive relevance of geometric structure relevant to EG in the semiclassical limit by doing away with the interior region of the black hole spacetime, as an effective description at and beyond the horizon relative to the choice of exterior weak-field observer at all moments in their proper time. 

But it also might not be! The treatment of the fauxrizon in the opening paragraph of the subsection --- as something that simply disappears in the semiclassical limit --- is telling. Namely: why should its disappearance in the semiclassical limit \emph{preclude} our alternatively representing the fuzzball, semiclassically from far away, as extending well into an interior? A fairly common operationalist stance in the foundations of GR would even seem to counsel the opposite: that, in having recovered in the semiclassical limit a description of an observer as located in an exterior black hole spacetime, one may then proceed to freely choose between distinct descriptions of that same observer's surrounding circumstance: as either terminating outside a horizon or continuing on through it. Either choice is descriptively adequate, for the sake of our representing the astrophysical black hole --- now understood to be a fuzzball --- \emph{as a spacetime in the semiclassical limit} (relative to the exterior weak-field observer).

In other words, the paradox remains so long as the operationalist stance familiar from the classical theory GR continues to be viable in a semiclassical modeling context. If nothing precludes a description of the astrophysical black hole relative to the exterior weak-field observer in the semiclassical theory as extending through a horizon into a black hole interior, then the physical reasoning underpinning EG persists in exactly the case where one has freely chosen to consider the observer's surrounding exterior black hole spacetime as itself extended through a distant horizon and into a black hole interior. Consequently, adequate representations of the astrophysical black hole --- now understood to be a fuzzball --- \emph{as a spacetime in the semiclassical limit} can exhibit the peculiar global geometric features associated with EG, in contradiction with what LM and GH entail. (Those peculiar global geometric features would, perhaps, indicate that there is something \emph{misleading about} such a choice of semiclassical description --- but it would still be appropriate \emph{as} one such description, and so the paradox would formally persist.)

\subsection{What constitutes a (quantum gravity) solution to semiclassical paradox}\label{sec:successortheory}

Where we have wound up so far is that it fails to suffice to solve the paradox that the fuzzball construction be free of conceptual difficulties reminiscent of the evaporation paradox, now formulated with regards to astrophysical modeling applications of string theory. Instead, to solve the paradox (disregarding loopholes, cf. footnote \ref{fn:loophole}), something about the stringy description of the astrophysical black hole as a fuzzball must supply us with cause to \emph{necessitate abandoning} EG: grounding a claim that we are, ultimately, not free in the application of the semiclassical theory to choose to represent a fuzzball, as considered in the semiclassical limit relative to an exterior weak-field observer, by means of a spacetime that includes a horizon and black hole interior. In other words, something about the demarcation of effective spacetime from the fuzzball's quantum hair relative to the exterior weak-field observer, i.e. the fauxrizon itself, must force our hand in describing a fuzzball, semiclassically, just in terms of an exterior black hole spacetime. (This, even while we simultaneously understand the fauxrizon itself to disappear in that same limit!) Something about the fauxrizon must render the complete black hole spacetime in the semiclassical limit, relative to any such observer, \emph{inadequate} as a description of that observer's physical circumstance.

Unfortunately, it is obscure what would demonstrate such a conclusion. But I can point to a toy version of the same punchline, in the hopes of spurring further ideas along these lines. What I have in mind is a case where detailed inspection of a quantized successor theory (with native structures that disappear in the classical limit) precludes from descriptive relevance certain states of affairs in the classical limit, which might otherwise have been kinematically plausible as part of a classical predecessor theory. This is an ongoing project by Benjamin \citeauthor{feintzeig2020classical} concerning systems of mechanics with finite degrees of freedom; here I will focus on the line of argument developed in just two recent articles \citep{feintzeig2020classical,feintzeig2020reductive}.

\citeauthor{feintzeig2020classical} draws on tools from quantization theory appropriate for systems with finite degrees of freedom, in order to offer a precise sense in which two closely related statements hold. First, the kinematic structure of a quantized theory of mechanics (characterized by canonical commutation relations) constrains the classical theory in the limit, which we may thereby regard as suitable for providing approximate descriptions of a quantum system ultimately described by an application of the quantized theory \citep{feintzeig2020classical}. Hence, facts about a quantized theory may plausibly delimit the descriptive scope of a classical theory applied to any such quantum systems. Second, requiring that the quantized theory accounts for the descriptive successes of its classical \emph{predecessor theory} constrains the activity of constructing the quantized theory \emph{as a successor theory}, given an interpretation of the classical one \citep{feintzeig2020reductive}. 

The relevance of the explicit toy example to the present case is not found in the technical procedures developed there, for the case of classical limit descriptions of quantum mechanical systems --- just so, I have neglected to discuss any of the technical work that forms the backbone of \citeauthor{feintzeig2020classical}'s project. Rather, what is important is the philosophical thesis staked out by means of that technical work, concerning intertheoretic relations as a \emph{sophisticated, non-trivial contribution} to the full development of a theory, particularly in light of that theory's genealogy: its being intended as successor to some other interpreted, empirically apt predecessor that we take to be descriptive in application to (many of) the same physical systems. 

Adopting Feintzeig's perspective to the present case, my suggestion is that where certain paradoxes might otherwise spoil the descriptive adequacy of the (semi)classical predecessor theory in certain modeling applications on a standard interpretation, we might ultimately be constrained in our writing down the new quantized theory \emph{so that} those paradoxes are circumvented in the appropriate limit. That is to say, in the present case, one might look forward to i) the string theory description of the astrophysical black hole as a fuzzball constraining adequate descriptions of the same system in a semiclassical limit (such as forcing the global spacetime property of extendibility there, in virtue of facts about the fauxrizon --- cf. footnote \ref{fnfirewarll}) and ii) our having developed string theory \emph{so that the paradox will, in that limit, turn out to be solved}, accordingly.

That fauxrizons in a quantum gravity description of an astrophysical black hole as a fuzzball may intertheoretically constrain our semiclassically modeling them is an interesting proposal, for at least two reasons. First, it sharpens our attention in quantum gravity research toward the ongoing status of global spacetime structure in GR, now construed as a subject to be understood in terms of a semiclassical limit. As elaborated by \citet{manchak2018information}, building on themes familiar from \citet{earman1995bangs}, it is a mistake to think of the causal pathology involved with full evaporation as merely amounting to there being some `missing point', at which the metric cannot be smoothly, locally extended --- i.e. a `naked singularity' (see footnote \ref{fnNaked}). Indeed, the present discussion might suggest that we are learning about that naked singularity's ultimately stringy structure, so that what we ordinarily have cause to think of as a formal, quasi-local breakdown of the theory in the semiclassical limit is, in fact, a kind of fundamental physical --- namely, stringy --- system in itself. One might even wonder, along these lines, whether the global spacetime description featuring a naked singularity \emph{just is} a mistaken semiclassical treatment of what is, ultimately, the dynamical production of a stringy black hole remnant, relative to a choice of exterior weak-field observer. This is reminiscent of `Attitude 2a' in the recent taxonomy provided by \citet{crowther2021four}, where singularities in GR are regarded as physically significant, in the sense of being informative of \emph{new} physics.

Second, and along similar lines, it is curious that in order for the fuzzball proposal to give us cause to necessitate abandoning EG, it evidently must be in virtue of the fauxrizon --- the quasi-local fade-out of an effective description of spacetime, which quarantines quantum hair associated with the black hole interior at what would otherwise be a horizon --- that EG is rendered false. The claim that, in a string theory approach to quantum gravity research, the fuzzball proposal ultimately solves the evaporation paradox would therefore seem to be a claim that some quasi-local, fundamental degrees of freedom in string theory wind up constraining global spacetime structure in the choices we make in our representation of that same physics in the semiclassical limit. In some sense, this is unsurprising: even before introducing back-reaction to motivate evaporation dynamics, as noted, Hawking radiation associated global states of affairs with the quasi-local production of modes along a horizon, which would otherwise in GR be unremarkable, locally, as a surface in the spacetime. But it is nonetheless tantalizing that local dynamics at the level of fundamental description are sought, which would intertheoretically constrain the specification of the \emph{boundary} in an observer-relative application of the limit theory (in the sense of constraints relevant to solving a local geometrodynamics in the bulk). Most discussions about intertheoretic relations in fundamental physics research have rather so far focused exclusively on the capacity of a fundamental theory to constrain effective descriptions of dynamics in the bulk (without attention to any changes in description of the boundary, nor to the role of the schematic of the observer carried through the intertheory relation). 

\section{Conclusion}

The purpose of this article was to elaborate on a natural suggestion that fauxrizons might ultimately solve the evaporation paradox in quantum gravity research, specifically in the case of the fuzzball proposal in string theory (or quantum gravity firewalls, more generally --- cf. footnote \ref{fnfirewarll}). As stressed throughout, it is not merely the fact that a stringy astrophysical black hole, modeled as a fuzzball, has a fauxrizon, which would imply that the proposal solves the evaporation paradox in a string theory approach to quantum gravity research. Rather, something about the fauxrizon must force our hands, in some or other way, in the recovery of the application of semiclassical gravity to the fuzzball in a suitable limit. Indeed, on this point, it is curious that we might hope to learn about a feature of global spacetime structure in an observer-relative application of semiclassical gravity at the limit, in virtue of quasi-local stringy dynamics (relative to the same observer). 

The focus here on descriptions of the boundary across intertheory relations, and in particular the emphasis on the role of the schematic of the observer in modeling on both sides, may be of some general significance in the foundations of quantum gravity research: informing our thinking about such other familiar quantum gravity modeling topics as Big Bang singularity resolution and the emergence of (perturbed, approximately uniformly expanding) spacetime in quantum cosmology. Meanwhile, the careful treatment throughout of what does and does not suffice to solve a paradox in current physics by means of further theorizing is, I believe, of consequence in a more general methodology or philosophy of science, which is chiefly concerned with the long-term fate of our current understanding in physics --- or (just as well) the continuity of our understanding in physics, across potentially radical theory change in the not-too-distant future. On this topic, the emphasis in \S\ref{sec:successortheory} on the sustained descriptive adequacy of applications of a predecessor theory in the wake of its intended successor is, I hope, clear: at least in some cases, there is a sharp, methodologically pertinent distinction between a claim to have discarded a paradox in the course of theoretical research (say, by moving to a new theory whose application to an old modeling target plausibly evades the paradox) and a claim to have solved it  (thereby learning something new).

\bibliographystyle{abbrvnat}
\bibliography{evaporationparadox}
\end{document}